\newcommand{\BibDirectory}{.}

\documentclass[aps,prl,twocolumn,twoside,bibnotes,preprintnumbers]{revtex4}

\usepackage{amsfonts,amssymb,amsmath}         
\usepackage{graphics,graphicx,epsfig}         

\newcommand{\ssc}[2][]{^{\vphantom\dagger{#1}}_{#2}}
\newcommand{\tr}[1][]{\mathop{{\mathrm{tr}}\ssc{#1}}}
\newcommand{\ket}[2][]{\left|{#2}\right\rangle\mskip-5mu\ssc{#1}}

\begin{document}


\title{Universal quantum interfaces}

\author{Seth Lloyd}
\email[]{slloyd@mit.edu}

\affiliation{d'Arbeloff Laboratory for Information Systems and Technology,
             Massachusetts Institute of Technology,
             Cambridge, MA 02139, USA}

\author{Andrew J. Landahl}
\email[]{alandahl@mit.edu}
\affiliation{Center for Bits and Atoms,
             Massachusetts Institute of Technology,
             Cambridge, MA 02139, USA}
\affiliation{HP Labs,
             Palo Alto, CA 94304-1126, USA}

\author{Jean-Jacques E. Slotine}
\email[]{jjs@mit.edu}

\affiliation{Nonlinear Systems Laboratory,
             Massachusetts Institute of Technology,
             Cambridge, MA 02139, USA}


\date[]{10 Mar 2003}
%


\begin{abstract}

To observe or control a quantum system, one must interact with it via an
interface.  This letter exhibits simple universal quantum interfaces---quantum
input/output ports consisting of a single two-state system or quantum
bit that interacts with the system to be observed or controlled.  It is shown
that under very general conditions the ability to observe and control the
quantum bit on its own implies the ability to observe and control the system
itself. The interface can also be used as a quantum communication channel, and
multiple quantum systems can be connected by interfaces to become an efficient
universal quantum computer. Experimental realizations are proposed, and
implications for controllability, observability, and quantum information
processing are explored.

\end{abstract}


\maketitle



A common problem in quantum control and quantum computation is that of building
up complex behaviors out of simple operations. For instance, considerable
effort has been devoted to investigating how to efficiently control the state
and the dynamics of complex quantum systems
\cite{geometric-control:Huang:1983a+,Lloyd:2000b,quantum-feedback-control:Wiseman:1993b+}.
Dual to the problem of controlling complex systems is that of observing them. 
Both controllers and observers are needed for feedback control of quantum
systems \cite{Lloyd:2000b,quantum-feedback-control:Wiseman:1993b+}.  In quantum
computation, quantum logic gates are simple local operations that can be
combined to manipulate quantum information in any desired way
\cite{any-gate-universal:Lloyd1995a+}. Quantum control and quantum computation
are fundamentally based on getting and processing information
\cite{Touchette:2000a}.

Geometric control theory has been used to show the universality of simple
quantum operations for performing coherent control
\cite{geometric-control:Huang:1983a+,Lloyd:2000b} and for quantum computation
\cite{any-gate-universal:Lloyd1995a+}.  In particular, almost any pair of
Hamiltonians that can be applied to a closed, finite-dimensional quantum
system render it controllable, and almost any quantum logic gate is universal
\cite{any-gate-universal:Lloyd1995a+}. Less attention has been paid to the
problem of observability; however, it is known that coherent controllability
of a quantum system combined with the ability to perform simple measurements
on it renders the system observable \cite{Lloyd:2001a}.  Specific examples of
systems that interact with a quantum system to control and observe it were
investigated in \cite{qcontroller-system:Janzing2002+}. Quantum feedback
control can be used to protect systems from disturbances
\cite{quantum-feedback-noise-suppression:Tombesi1995a+} and to engineer
open-system dynamics \cite{Lloyd:2001a}. Quantum error correction can be used
to protect quantum information from noise and decoherence
\cite{quantum-error-correction:Shor1995a+}. This Letter exhibits a simple
quantum device---a universal quantum interface, or UQI---that is able to
perform all these tasks simply and efficiently.  The universal quantum
interface consists of a single two-state quantum system, or quantum bit, that
couples to a Hamiltonian system to be controlled or observed via a fixed
Hamiltonian interaction.  The primary purpose of this Letter is to show that
by controlling and observing the quantum bit on its own, one can fully control
and observe the system to which it is coupled.

Consider a $d$-dimensional quantum system $S$ whose dynamics are described by
a Hamiltonian $H$. Consider a two-level system $Q$ coupled to $S$ via a fixed
Hamiltonian interaction $A\otimes \sigma_z$ where $A$ is an Hermitian operator
on $S$ and $\sigma_z$ is the $z$ Pauli matrix with eigenvectors $|+1\rangle$
corresponding to eigenvalue $+1$ and $|-1\rangle$ corresponding to eigenvalue
$-1$. Assume that we can both make measurements on $Q$ in this basis, and
apply Hamiltonians $\gamma \sigma$ to $Q$, where $\sigma$ is an arbitrary
Pauli matrix and $\gamma$ is a real control parameter.  That is, taken on its
own, $Q$ is controllable and observable (the ability to measure with respect
to one basis combined with the ability to perform arbitrary rotations
translates into the ability to measure with respect to any basis).

It can immediately be shown that in the absence of environmental interactions
the system is generically coherently controllable.  As long as $H$ and $A$ are
not related by some symmetry, the algebra generated by $\{H+A\otimes \sigma_z,
\gamma\sigma\}$ is the whole algebra of Hermitian matrices for $S$ and $Q$
taken together. As a result, by the usual constructions of geometric control
theory \cite{geometric-control:Huang:1983a+}, one can perform arbitrary Hamiltonian
transformations of the system and qubit by turning on and off various
$\sigma$s. One such Hamiltonian transformation is an arbitrary Hamiltonian
transformation on the system on its own, so the system is coherently
controllable.

Now turn to observability.  Since by controlling the qubit on its own we can
engineer any desired Hamiltonian transformation of the system and qubit
together, we can apply any evolution of the form $e^{-iG\otimes \sigma_x t}$,
where $G$ is an arbitrary Hermitian operator on $S$ and $\sigma_x$ is the
$x$-Pauli matrix on $Q$. Prepare the interface in the state
$\ket{+1}$ (e.g., by measuring the qubit and rotating it to $\ket{+1}$), apply
this evolution, and measure $Q$ in the  $\{\ket{+1},\ket{-1}\}$ basis. As a
result of this preparation, evolution, and measurement, the system state
evolves from $\rho_S(0)$ into either $\rho_S^+ = \cos(\gamma t
G)\rho_S(0)\cos(\gamma t G)$ or $\rho_S^- = \sin(\gamma t
G)\rho_S(0)\sin(\gamma t G)$, with probabilities $p_+ = \tr \cos^2(\gamma t
G)\rho_S(0)$ and $p_- = \tr \sin^2(\gamma t G)\rho_S(0)$ respectively.  In
other words, this procedure effects the generalized ``Yes-No" measurement on
$S$ having Hermitian Kraus operators $\cos(\gamma t G)$, $\sin (\gamma t G)$. 
This is the form of the most general minimally-disturbing two-outcome
measurement on $S$ \cite{Barnum:1998a}.  In \cite{Lloyd:2001a}, it is shown
how one can perform any desired generalized measurement corresponding to Kraus
operators $\{ A_{k} \}$ by making a series of such two-outcome measurements. 
So by the construction outlined above, where the results of the two-outcome
measurements are copied to classical memory, $Q$ can effect an arbitrary
generalized measurement on $S$ and is therefore a full semiclassical observer
for $S$ \cite{Lloyd:2000b}.

\begin{figure}
\includegraphics[width=2.3in]{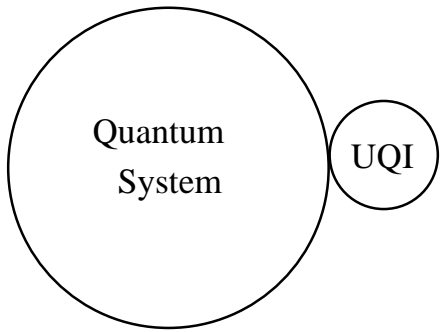}
\caption{A universal quantum interface attaches itself to a system
with Hamiltonian $H$ via an interaction $A\otimes\sigma_z$.  By
measuring and manipulating the single qubit of the interface, one
can control and observe the quantum system in any desired way.}
\label{fig:uqi_controller}
\end{figure}

Generalized measurements and generalized open-system transformations
are closely related.  By making a generalized measurement and ignoring
the outcomes one effects the open-system transformation
$\rho_S(0) \rightarrow \sum_k A_k \rho_S(0) A^\dagger_k$.
So our universal quantum interface $Q$ is not only a full semiclassical
observer for $S$, but also a universal controller capable of performing any
desired completely positive linear trace-preserving map on $S$
\cite{Kraus:1983a} (see Fig.~\ref{fig:uqi_controller}).

\begin{figure}
\includegraphics[width=2.3in]{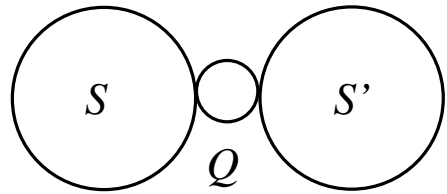}
\caption{A universal quantum interface that interacts with two systems
can serve as a quantum communication channel, mediating the flow of
information between the two systems.}
\label{fig:uqi_comm_channel}
\end{figure}

True to its name, the universal quantum interface can also act as
a quantum communication channel between two quantum systems, $S$
and $S'$.  Let $Q$ be coupled to $S$ with a coupling $A\otimes\sigma_z$
and to $S'$ with a coupling $A'\otimes \sigma_z$.  As long as the
algebras generated by $\{H,A\}$ and by $\{H',A'\}$ close only on the full
algebras for the two systems on their own, then the algebra generated by
$\{H+H'+ A\otimes\sigma_z + A'\otimes \sigma_z$, $\gamma\sigma\}$
closes on the full algebra for the two systems together with $Q$.
Consequently, $Q$ can be used to shuttle quantum information from
$S$ to $S'$ and {\it vice versa} (see Fig.~\ref{fig:uqi_comm_channel}).

The ability of quantum interfaces to perform communication tasks
as well as coherent quantum information manipulation and measurement
allows one to envisage a quantum control system, including sensors,
controllers, and actuators, constructed of quantum systems linked
via quantum interfaces, or even constructed entirely of quantum interfaces
in series and parallel.  Such quantum control systems could
effect either coherent or incoherent quantum feedback
\cite{Lloyd:2000b,quantum-feedback-control:Wiseman:1993b+}.

The universal quantum interface can control a quantum system, observe it, and
shuttle quantum information between systems.  How efficiently can it perform
these tasks?  Here we can use an argument based on the Solovay-Kitaev theorem
\cite{Nielsen:2000a}.  The transformations on the system and interface
correspond to a time-dependent Hamiltonian $H+A\otimes\sigma_z +
\gamma(t)\sigma(t)$. Arbitrary unitary transformations on system and interface
can be built up this way. Let $\tau$ be the characteristic time that it takes
to build up two unitary transformations $U$, $U'$ that differ significantly
from each other (i.e., $\tr U^\dagger U' \ll d$). Assuming that the unitary
transformations that can be built up over times much greater than $\tau$ are
distributed essentially uniformly over the space of all unitary
transformations, one sees that in time $t$ one can perform an arbitrary
control or observation on a $d$-dimensional quantum system to an accuracy
proportional to $e^{-t/\tau d^2}$.  To obtain exponential accuracy requires
time of $O(d^2)$.

Of course, some control tasks and observations can be performed in less time. 
In the case of quantum computing, we are interested in transformations of $n$
qubits, so that the dimension of the Hilbert space is $d=2^n$. A generic
transformation can be built up out of $O(2^{2n})$ quantum logic gates.  But
some computations (Shor's algorithm \cite{Shor:1994a}, and quantum simulation
\cite{Lloyd:1996a}, for example) can be performed in time polynomial in $n$,
i.e., polylogarithmic in $d$.

A universal quantum interface can effect any desired transformation on the
system to which it is connected, including quantum logic transformations.  But
if the system to which it is connected is high-dimensional, e.g., $d=2^n$, the
interface cannot necessarily effect those transformations efficiently.  In
particular, a desired quantum logic operation could take time $O(2^{2n})$ to
effect. The general condition on $H$ and $A$ under which it is possible to
perform quantum computation efficiently on a $d=2^n$ dimensional system is an
open question.

\begin{figure}
\includegraphics[width=2.3in]{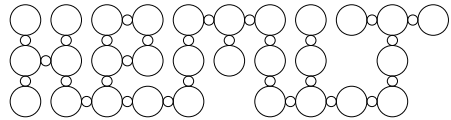}
\caption{A set of quantum interfaces connecting low-dimensional systems
makes up a quantum computer, capable of performing quantum logic operations
and shuttling information between any two subsystems.}
\label{fig:uqi_network}
\end{figure}

If one uses quantum interfaces to control and connect a number of quantum
systems, however, one can in general perform efficient universal quantum
computation. A specific architecture in which universal quantum interfaces can
be used to perform universal quantum computation is one in which $n$
small-dimensional systems are coupled together via quantum interfaces as
described above (see Fig.~\ref{fig:uqi_network}). Any set of pairwise
couplings between systems that forms a connected graph now allows efficient
universal quantum computation as follows.

First, consider the problem of performing coherent quantum logic operations on
the coupled systems. Prepare the interfaces in the state $|+1\rangle$ by
measuring them. Each interface is now in an eigenstate of the Hamiltonians
$H_j+A_j\otimes \sigma_z$ that couples it to its connecting systems.  As a
result, the systems are all effectively uncoupled and evolve by renormalized
versions of their respective Hamiltonians: the $j$th system evolves via the
Hamiltonian $H_j + A_j$.  By coherently controlling the interface between
the $j$th and $k$th system, one can effect an arbitrary coherent
transformation of these two systems together, returning the interface to the
state $|+1\rangle$.  That is, one can perform any desired quantum logic
transformation on any two systems that are connected by an interface.  While
this quantum logic transformation takes place, the other systems evolve in an
uncoupled fashion via known Hamiltonians.

Since the graph that describes the interfaces is fully connected, quantum
information can be moved at will throughout the set of coupled systems by
sequential pairwise couplings intermediated by the interfaces.  The maximum
number of pairwise operations required to bring any two qubits into adjacent
systems is $O(n)$.  Arbitrary quantum logic transformations can be performed
on systems in a pairwise fashion.  As a result, any desired quantum logic
circuit of $N$ logic gates can be built up using no more than $O(d^2nN)$
pairwise operations, where $d$ is the typical dimension of a subsystem.  If
the systems are qubits then the quantum logic circuit can be built up in
$O(nN)$ operations.  For example, the coupled systems could themselves be
quantum interfaces, so that an entire quantum computer could be constructed
from interfaces alone.

State preparation and measurement can be accomplished in a similar
fashion.  By manipulating and measuring a given interface, while
keeping the other interfaces `turned off' via the decoupling
procedure given above, one can perform any desired generalized
measurement on the systems to which that interface is coupled.
This procedure allows one both to prepare and to measure the state
of those systems.  Since state preparation, coherent quantum logic
operations, and measurement can all be accomplished efficiently,
the set of systems coupled by universal interfaces can perform
universal quantum computation.

Universal quantum interfaces are simple systems that can be used to perform
arbitrary quantum operations---control, observation, and computation---on
quantum systems.  Note that the derivations above depend on the fact that the
systems to be controlled or observed are closed apart from the interactions
with their interfaces. If the systems to be controlled or observed are open to
the environment, as all systems are to a greater or lesser degree (`no quantum
system is an island entire unto itself'), then only those operations which can
be performed efficiently within the system's decoherence time can actually be
effected.  An interesting open question for further research is the degree to
which quantum interfaces can be used to protect quantum systems and
effectively decouple them from their environment via the use of symmetries
\cite{decoherence-free-subspaces:Zanardi1997a+}, bang-bang techniques
\cite{Viola:1998a}, or analogs of quantum error correcting codes
\cite{quantum-error-correction:Shor1995a+}.

The straightforward requirements for universality allow many candidates for
quantum interfaces.   Many systems that are frequently used to couple to
quantum systems are universal quantum interfaces.  For example, a mode of the
electromagnetic field that couples to an optical cavity can be used to control
and observe the contents of the cavity, as in quantum computing using cavity
quantum electrodynamics \cite{Doherty:2000a}.  In an ion trap, the internal
and vibrational states of the ions could be controlled and observed using just
one ion in the trap (for example, an ion of a different species from the other
ions in the trap \cite{Blinov:2001a}). In general, a single optically active
site on a molecule, e.g., one held in optical tweezers to minimize coupling to
the environment, could be used to control and observe the quantum states of
the molecule. If the electronic and hyperfine states of the atoms in the
molecule can be addressed either individually or in parallel, such a molecule
addressed via an optical quantum interface is a good model for quantum
computation.  In liquid state NMR, it is possible to control and observe the
state of the nuclear spins in a molecule by observing just one nuclear spin on
the molecule while using coherent control to shuttle quantum information from
the spins to be observed to the observed spin \cite{nmr-qubit:Cory1998+}. In
coherent superconducting circuits, for example ones made up of several coupled
charge or flux qubits, the state of the entire circuit can in general be
coherently controlled and observed simply by controlling and observing a
single qubit, which could be specially designed for this purpose
\cite{superconducting-qubit:Mooij1999}.

Universal quantum interfaces are devices that can be used to
control and observe a quantum system in any desired fashion.
Because of their simple nature, universal quantum interfaces are
considerably easier to exhibit experimentally than is a universal
quantum computer.  Indeed, existing interfaces with cavity QED, ion-trap,
and NMR systems are already universal.  Networks of quantum interfaces
can be used to perform arbitrarily difficult quantum control tasks
in principle, including full-blown quantum computation.  In practice,
complicated quantum information processing tasks
involving many quantum interfaces are of the same order of difficulty
to perform as quantum computation.  Open questions include problems
of efficiency, networkability, and interfaces with quantum
systems that interact strongly with their environment.


\acknowledgments

This work was supported by the HP/MIT collaboration.




\bibliography{\BibDirectory/landahl_thesis}

\end{document}